\documentclass[12pt,a4paper]{article}
\usepackage{cite,graphicx,amsmath,amssymb,colordvi,color}

\def\beq{\begin{equation}}
\def\eeq{\end{equation}}
\def\bea{\begin{eqnarray}}
\def\eea{\end{eqnarray}}
\def\beaa{\begin{eqnarray*}}
\def\eeaa{\end{eqnarray*}}

\def\putunder#1#2{\mathrel{
\setbox0=\hbox{#1}\setbox1=\hbox{\scriptsize #2} \dimen0=-0.5\wd0
\advance\dimen0 by -0.5\wd1 \dimen1=0.5\wd0 \advance\dimen1 by -0.5\wd1
\hbox{\box0\kern\dimen0%
\vbox to 0pt {\hbox{\lower 0.7em \box1}\vss}%
\kern\dimen1} }}

\newcommand{\gsim}{\lower.7ex\hbox{$\;\stackrel{\textstyle>}{\sim}\;$}}
\newcommand{\lsim}{\lower.7ex\hbox{$\;\stackrel{\textstyle<}{\sim}\;$}}
\newcommand{\be}{\begin{equation}}
\newcommand{\e}{\end{equation}}

\begin{document}

\thispagestyle{empty}
\vspace*{.5cm}
\noindent
DESY 05-118\hspace*{\fill} July 7, 2005\\
\vspace*{1.6cm}

\begin{center}
{\Large\bf Eternal Inflation with $\alpha'$-Corrections}
\\[2.0cm]
{\large A. Westphal}\\[.5cm]
{\it Deutsches Elektronen-Synchrotron, Notkestrasse 85, D-22603 Hamburg,
Germany}
\\[3cm]

{\bf Abstract}\end{center} \noindent Higher-order
$\alpha'$-corrections are a generic feature of type IIB string
compactifications. In KKLT-like models of moduli stabilization
they provide a mechanism of breaking the no-scale structure of the
volume modulus. We present a model of inflation driven by the
volume modulus of flux compactifications of the type IIB
superstring. Using the effects of gaugino condensation on
D7-branes and perturbative $\alpha'$-corrections the volume
modulus can be stabilized in a scalar potential which
simultaneously contains saddle points providing slow-roll
inflation with about $130$ $e$-foldings. We can accommodate the
3-year WMAP data with a spectral index of density fluctuations
$n_s=0.93$. Our model allows for eternal inflation providing the
initial conditions of slow-roll inflation.

\newpage

\section{Introduction}

String theory at present is the only candidate for a unified quantum theory of
all interactions that simultaneously provides for a UV-finite description of
quantum gravity. However, there is rich internal structure already in 10
dimensions and the tremendously large number of possible compactifications to
4d (roughly $10^{500}$ according to a recent estimate~\cite{dougl,sussk}).
Thus, we face the formidable task of constructing realistic 4d string vacua
that come as close as possible to the structures of the Standard Model. One
pressing issue is removing the massless compactification moduli from the low
energy spectrum of a given string vacuum. Recently, more general
compactification manifolds characterized by the presence of background
fluxes~\cite{GKP,CBachas,PolStrom,GVW,Michelson,
DasSeRa,TaylVaf,Vafa,Mayr,GSS,KlebStrass,Curio2,CKLT,HaaLou,BB,DallAgata,KaScTr,silver,acharya,dlust}
of the higher $p$-form field strengths in string theory have been studied in
this context. Such flux compactifications can stabilize the dilaton and the
complex structure moduli in type IIB string theory. Non-perturbative effects
such as the presence of Dp-branes~\cite{Verl} and gaugino condensation were
then used by KKLT~\cite{kklt} to stabilize the remaining K\"ahler moduli in
such type IIB flux compactifications (for related earlier work in heterotic
M-theory see~\cite{Curio1}). Simultaneously these vacua allow for SUSY breaking
and thus the appearance of metastable $dS_4$-minima with a small positive
cosmological constant fine-tuned in discrete steps. KKLT~\cite{kklt} used the
SUSY breaking effects of an anti-D3-brane to achieve this. Alternatively the
effect of D-terms on D7-branes have been considered in this
context~\cite{bkqu}.

Concerning KKLT inspired setups like those mentioned above we may now ask which
of the ingredients used there is least controlled with respect to the
constraints of perturbativity and negligible backreactions. Clearly, such a
question arises with the use of anti-D3-branes as uplifts for given
volume-stabilizing AdS minima. The presence of either D3-branes or
anti-D3-branes by themselves does not pose a problem. Each kind viewed for
itself is a BPS state that preserves half of the original ${\cal N}=8$
supersymmetries in 4d (${\cal N}=2$ in 10d), which, in turn, can be arranged to
contain the 2 supersymmetries preserved by the Calabi-Yau compactification.
However, an anti-D3-brane in the presence of a compact geometry with D3-branes
is non-BPS with respect to the supersymmetries preserved by the BPS condition
of the D3-branes. Thus, it breaks SUSY, and it is not clear whether this SUSY
breaking is explicit or has a description in terms of F-term or D-term
breaking. If anti-D3-branes break SUSY explicitly, the use of the supergravity
approximation to calculate the effect on the scalar potential may be
questionable. Replacing the anti-D3-branes by D-terms on D7-branes~\cite{bkqu}
is a way to alleviate this problem because this way of SUSY breaking has a manifestly
supersymmetric description.

In view of these difficulties it is appealing that there are further
possibilities to provide uplifting effects by means of perturbative
$\alpha^{\prime}$-corrections~\cite{bbhl} (for earlier results
see~\cite{bbhlpre}) in the type IIB superstring. KKLT have argued that these
higher-order corrections in the string tension are not relevant in the large
volume limit~\cite{kklt}. The non-perturbative effects invoked by KKLT vanish
exponentially fast in this limit. In contrast, the perturbative corrections
usually depend on a power of the volume. This motivates the discussion of these
effects as an alternative to anti-D3-branes. The $\alpha'$-corrections have
recently been used to provide a realization of the simplest KKLT $dS$-vacua
without using anti D3-branes as the source of SUSY
breaking~\cite{Brama,Bobk,Brama2,Conlon}. Here we will show that in combination
with racetrack superpotentials these stringy corrections can provide also for
slow-roll inflation driven by the KKLT volume modulus. Inflation in string
theory has been studied recently by, e.g, using the position of
D3-branes~\cite{kklmmt,braneinf} or a condensing D-brane tachyon~\cite{tachyon}
as the inflaton field (for recent attempts to cure the $\eta$-problem of
supergravity in such brane inflation models see e.g.~\cite{BBKraus,brax}) or
tuning the original KKLT potential for the KKLT volume modulus $T$ by extending
the superpotential used there to the racetrack type~\cite{pill,ross}. The
general evolution of the $T$-modulus was studied in~\cite{modevol} for the KKLT
case~\cite{kklt} and the modified Kallosh-Linde model~\cite{kalllind}.

The paper is organized as follows. Section~\ref{alpha} summarizes
known $\alpha'$-corrections in type IIB superstring theory and
provides a short discussion of the scalar potential generated by
these corrections. Section~\ref{nokklt} discusses the uplifting
potential provided by $\alpha'$-corrections in terms of a general
limiting case of the form of the uplifting contribution. We show
that the KKLT superpotential combined with one or two additive
uplifting contributions to the scalar potential cannot provide for
slow-roll inflation driven by the $T$-modulus. This result is used
afterwards in Sect.~\ref{Tinfl} to motivate the extension of the
KKLT case to a superpotential of the racetrack type. Once we
combine a racetrack superpotential with the $\alpha'$-corrections
the $T$-modulus acquires a scalar potential which stabilizes this
field at a weakly $dS$-minimum. Simultaneously this scalar
potential contains saddle points which are sufficiently flat to
provide for more than $130$ $e$-foldings of slow-roll inflation
driven by the $T$-modulus. Section~\ref{rescale} discusses
important rescaling properties of the setup. In
Section~\ref{ExpCon} we construct a phenomenologically viable
model of $T$-modulus inflation along these lines which can
accommodate the 3-year WMAP data~\cite{WMAP} of the CMB radiation.
It yields primordial density fluctuations of the right magnitude
with a spectral index of these fluctuations $n_s\approx 0.93$. In
Section~\ref{infla3} we check our numerical results within the
analytical treatment of inflation on a generic saddle point. We
find that the inflationary saddle points of the model allow for
eternal topological inflation. Finally, we summarize our results
in the Conclusion.

\section{$\alpha'$-corrections}\label{alpha}

Higher-order $\alpha^{\prime}$-corrections which usually lift the no-scale
structure of the K\"ahler potential of the volume modulus (and generate 1-loop
corrections to the gauge kinetic functions) are not known in general. However,
there is one known perturbative correction~\cite{bbhl} given by a
higher-derivative curvature interaction on Calabi-Yau threefolds of
non-vanishing Euler number $\chi$. Its relevant bosonic part is given as \bea
S_{\textrm{IIB}}&=&\frac{1}{2\kappa_{10}^2}\int d^{10}x\sqrt{-g_{\textrm
s}}\;e^{-2\phi}\left[R_{\textrm
s}+4\left(\partial\phi\right)^2+\left.\alpha^{\prime}\right.^3\frac{\zeta(3)}{3\cdot
2^{11}}\;J_0\right]\;\;.\label{aprim}\eea Here $J_0$ denotes the
higher-derivative interaction~\cite{bbhl}\[J_0=\left(t^{M_1N_1\cdots M_4N_4}
t_{M_1^{\prime}N_1^{\prime}\cdots M_4^{\prime}N_4^{\prime}}
+\frac{1}{8}\;\epsilon^{AB M_1N_1\cdots M_4N_4}\epsilon_{AB
M_1^{\prime}N_1^{\prime}\cdots M_4^{\prime}N_4^{\prime}}\right)
\left.R^{M_1^{\prime}N_1^{\prime}}\right._{M_1N_1}\cdots
\left.R^{M_4^{\prime}N_4^{\prime}}\right._{M_4N_4}\] which after Calabi-Yau
compactification to 4d yields a correction to the K\"ahler potential of the
volume modulus $T$~\cite{bbhl} \bea K&=&-2\cdot\ln\left({\cal
V}+\frac{1}{2}\;\hat{\xi}\right)\;\;,\;\hat{\xi}=\xi
e^{-3\phi/2}\;\;,\;\xi=-\frac{1}{2}\;\zeta(3)\chi\nonumber\\
&=&\underbrace{-3\cdot\ln\left(T+\bar{T}\right)}_{K^{(0)}}
-2\cdot\ln\left(1+\frac{\hat{\xi}}{2({2\;\textrm{Re}}\;T)^{3/2}}\right)\;\;.\label{dK}\eea
Here the volume modulus $T$ is related to the Calabi-Yau volume ${\cal V}$ as
${\cal V}=(T+\bar{T})^{3/2}$ (see, e.g.,~\cite{Grimm})\footnote{Here ${\cal V}$
is defined in the Einstein frame~\cite{bbhl}.}. $\chi$ denotes the Euler number
of the Calabi-Yau under consideration which can be of both signs and in its
absolute value can be at least as large as 2592~\cite{huschi}. From the general
expression for the scalar potential in 4d ${\cal N}=1$ supergravity the
potential for the $T$-modulus is \beq V(T)=e^K\left(K^{T\bar{T}}D_T W
D_{\bar{T}}\bar{W}-3\left|W\right|^2\right)\;\;.\label{VsugT}\eeq This leads to
a correction to the scalar potential of $T$ which to ${\cal
O}(\left.\alpha^{\prime}\right.^3)$ reads~\cite{bbhl} \beq \delta
V=-\frac{\hat{\xi}}{({2\;\textrm{Re}}\;T)^{3/2}}\;V_{\textrm{tree}}
+\frac{3}{8}\;e^{K^{(0)}}\frac{\hat{\xi}}{({2\;\textrm{Re}}\;T)^{3/2}}
\;\left|W+(\tau-\bar{\tau})\;\tilde{D}_{\tau}W\right|^2\label{alphcorr}\eeq
where $\tilde{D}_{\tau}W=\partial_{\tau}W+W\partial_{\tau}K^{(0)}$.
$V_{\textrm{tree}}$ denotes the full scalar potential for the volume modulus
$T$ except the effects of the $\alpha^{\prime}$-correction under discussion.

This correction, which breaks the no-scale structure of the K\"ahler potential
of the volume modulus, can be used as a replacement for the anti-D3-brane or
D-terms on D7-branes to provide the uplift necessary for realizing the KKLT
mechanism. Combining the KKLT ansatz for the superpotential \beq W(T)=W_0+A
e^{-aT}\label{WT2}\eeq with the $\alpha'$-correction is sufficient to realize
de Sitter vacua with all the moduli stabilized~\cite{Brama,Bobk,Brama2,Conlon}.
We can show now that a combination of the mechanism of uplifting by
$\alpha'$-corrections with a racetrack superpotential generates $dS$-minima
with full moduli stabilization. Simultaneously, the same potential contains
regions where $T$-modulus inflation with roll-off into the desired $dS$-minima
is realized. There is no $\eta$ problem in this setup because the leading order
K\"ahler potential of the volume modulus is of the no-scale type.

\section{Absence of $T$-modulus inflation in KKLT}\label{nokklt}

Before analyzing the setup sketched at the end of the last Section, we should
clarify why the original KKLT setup with just the superpotential
eq.~\eqref{WT2} and one uplifting correction $\delta V$ does not allow
$T$-modulus inflation. For this purpose, note that the types of uplift
considered so far can be written as \beq \delta
V=\frac{D}{X^{\alpha}}\;\;.\label{Vup1}\eeq Here we use that we write the
scalar component of the chiral superfield $T$ as $T|=X+iY$. Strictly speaking,
the above $\alpha'$-correction behaves as a mixture of additive and
multiplicative corrections. However, from the general form of the potential it
is clear that the above $\alpha'$-correction in the vicinity of the maximum can
be written locally in the same additive form \beq \delta
V=\frac{D}{X^{3/2}}\;,\;\;D=\left.\frac{\hat{\xi}}{2\sqrt{2}}\,\left(-V_{\textrm{tree}}+
\frac{3}{8}\,e^{K^{(0)}}\left|W+(\tau-\bar{\tau})\;\tilde{D}_{\tau}W\right|^2\right)
\right|_{T=T_{\textrm{max}}}\;\;.\label{alphaupl}\eeq

Thus, we may consider the following general setup: take the superpotential
Eq.~\eqref{WT2} to fix the $T$-modulus after the flux part $W_0$ has fixed all
the non-K\"ahler moduli. Add one uplifting term Eq.~\eqref{Vup1} with
$\alpha>0$ being general. Such a setup generically generates a maximum in the
$X$-direction separating the $dS$-minimum from infinity. Since this maximum
simultaneously forms a minimum in the $Y$-direction, we have the situation that
inflation would have to start from a saddle point with direction towards the
$dS$-minimum. For this purpose, two ingredients are necessary: firstly, a
definition of the slow-roll parameters for a scalar field with a
non-canonically normalized kinetic term. Secondly, an analysis of the scalar
potential's stationary points with respect to whether slow-roll can be
satisfied on the saddle or not.

The equations of motion for non-canonically normalized scalar
fields~\cite{klps,bcsq,kallpro,bgh} read \beq
\ddot{\phi}^l+3H\dot{\phi}^l+\Gamma^l_{ij}\dot{\phi}^i\dot{\phi}^j+G^{lk}\,\frac{\partial
V}{\partial\phi^k}=0\;,\;\;\Gamma^l_{ij}=-\frac{1}{2}\,G^{lk}\frac{\partial
G^{ij}}{\partial\phi^k}\;\;.\label{eomnc}\eeq For the $T$-modulus this implies
\beq G_{T\bar{T}}=K_{T\bar{T}}=\frac{3}{4X^2}\;\Rightarrow\;{\cal
L}_{\textrm{kin}}=
\frac{3}{4X^2}\,(\partial_{\mu}X\partial^{\mu}X+\partial_{\mu}Y\partial^{\mu}Y)\label{Lkin}\eeq
and thus the equations of motion become\bea
\ddot{X}+3H\dot{X}+\frac{1}{X}\,\dot{X}^2+\frac{2}{3}\,X^2\frac{\partial
V}{\partial X}&=&0\nonumber\\
\ddot{Y}+3H\dot{Y}+\frac{1}{X}\,\dot{Y}^2+\frac{2}{3}\,X^2\frac{\partial
V}{\partial Y}&=&0\;\;.\label{XYeom1}\eea The slow-roll parameters of, e.g.,
$X$ are thus given by \beq
\epsilon_X=\frac{X_{\textrm{max}}^2}{3}\left(\frac{V'}{V}\right)^2\;,\;
\eta_X=\frac{2 X_{\textrm{max}}^2}{3}\frac{V''}{V}\label{slow2}\eeq where $'$
denotes differentiation with respect to $X$.

The next step is to analyze the scalar potential. Including the uplift this
follows from Eq.~\eqref{VsugT} to be \beq
V(T)=\frac{1}{4X^2}\,\left\{2aA^2e^{-2aX}\left(1+\frac{1}{3}\,aX\right)+2aA W_0
e^{-aX}\cos(aY)\right\}+\frac{D}{X^{\alpha}}\label{VWT2}\;\;.\eeq The extrema
of this potential are determined by the conditions $\partial_X V=\partial_Y
V=0$. The $Y$-condition \beq\frac{\partial V}{\partial
Y}=0=-\,\frac{a^2A}{2X^2}\,e^{-2aX}W_0\sin(aY)\;\;\Rightarrow\;Y_{\textrm{extr}}=0\;\;
\textrm{for:}A W_0<0 \eeq implies that all extrema in $X$ are found along the
direction $Y=0$ with replications at $Y=\frac{2\pi n}{a}\;\forall
n\in\mathbb{Z}$. The extremal points are determined then by \bea \frac{\partial
V}{\partial
X}&=&0\nonumber\\
&\phantom{=}&\nonumber\\
\Leftrightarrow\;0&\approx&\frac{3\alpha
D}{aA}\,X^{2-\alpha}+\frac{3}{2}\,W_0\cdot\lambda(X)+A\lambda^2(X)
\;,\;\;\lambda(X)=aXe^{-aX}\label{kkltextr}\eea where we used the regime of
large volume \beq
aX\gg1\;\;,\;\;X\gg1\;\Rightarrow\;aA\gg\frac{A}{X}\gg\frac{W_0}{X}\;,\;
\frac{A}{X}\,e^{-aX}\ll\frac{W_0}{X}\label{constr}\eeq in order to trust the
use of the effective potential. Expanding the solutions to this quadratic
equation in $X^{2-\alpha}D/W_0^2\ll 1$ up to ${\cal
O}\Big(X_{\textrm{max}}^{2-\alpha}\frac{D}{W_0^2}\Big)$ leads to two extrema at
\beq
\frac{aX_{\textrm{max}}e^{-aX_{\textrm{max}}}}{X_{\textrm{max}}^{2-\alpha}}=-\,\frac{2\alpha
D}{a A W_0}\;,\;
aX_{\textrm{min}}e^{-aX_{\textrm{min}}}=-\,\frac{3W_0}{2A}\label{extr1}\eeq as
long as $A W_0<0$, which a posteriori justifies the use of this condition in
extremizing the potential in $Y$ above. Thus, the slow roll parameters of the
saddle are \beq
\epsilon_{X,\textrm{saddle}}=0\;\;,\;\;\eta_{X,\textrm{saddle}}=\frac{2}{3}\,X_{\textrm{max}}^2\,
\left.\frac{1}{V}\,\frac{\partial^2 V}{\partial
X^2}\right|_{X=X_{\textrm{max}},Y=0}=-\,\frac{2}{3}\,\alpha\,aX_{\textrm{max}}\;\;.
\label{kklteta1}\eeq Thus, we have $\eta\ll 1$ only if $\alpha\lsim 0.1$ (for
which no known realization exists) or $aX_{\textrm{max}}\lsim 1$, which
violates the large volume and perturbativity assumptions. Slow-roll inflation
with the $T$-modulus on the saddle point of this most simple class of KKLT-like
setups does not work.

Note that this condition corresponds to the fact that the single uplift $\delta
V$ already by itself has $\eta_{\delta V}=2/3\cdot X^2 \delta V''/\delta
V=2\alpha(1+\alpha)/3\ll 1$ for $\alpha\ll 1$. Thus $\delta V$ in general has
to behave nearly like a constant in order to generate a sufficiently flat
maximum of $V$.

We can extend this analysis immediately to the case of two additive uplifts
given by \beq \delta
V_2=\frac{D_1}{X^{\alpha_1}}+\,\frac{D_2}{X^{\alpha_2}}\;\;,\;\alpha_1,\alpha_2>0\;\;.
\label{twoupl}\eeq (Such a contribution might arise, e.g., if more than one
$\alpha'$-correction to the K\"ahler potential is included and locally written
in the above form, see Eq.~\eqref{alphaupl}. Unfortunately, none are known
besides the one of~\cite{bbhl}.) Without loss of generality we may assume
$\alpha_1<\alpha_2$. Then there are two cases.

In one situation we have both $D_1$ and $D_2$ positive implying that $\delta
V_2$ decreases strictly monotonically: $\delta V'_2<0$ and $\delta V''_2>0$
$\forall X>0$. This leads back to the above result with just one uplift and
thus to Eq.~\eqref{kklteta1} but with $\alpha$ replaced by some linear
combination $c_1\alpha_1+c_2\alpha_2\in[\alpha_1,\alpha_2]$ where $c_1+c_2=1$
with $0<c_1,c_2<1$.

The other and more interesting case is to have $D_1>0$ and $D_2<0$. Then
$D_2/X^{\alpha_2}$ is negative and strictly monotonically increasing for all
$X>0$ while $D_1/X^{\alpha_1}$ is positive and strictly monotonically
decreasing. Further, since we assumed $\alpha_1<\alpha_2$ we have $\lim_{x\to
0}\delta V=-\infty$. Therefore $\delta V_2$ has exactly one zero and one global
maximum within $(0,\infty)$. At the maximum $\epsilon_{\delta V_2}^{\rm
max}=0$. As we noted above $\delta V_2$ has to behave nearly like a constant in
order to provide a sufficiently flat maximum of $V$. This is realized close to
the maximum of $\delta V_2$ if we tune $\eta_{\delta V_2}^{\rm max}\ll 1$.
Using $X_{\rm max}$ determined by $\delta V'_2(X_{\rm max})=0$ we arrive at
\beq\eta_{\delta V_2}^{\rm max}=-\,\frac{2}{3}\cdot\alpha_1\alpha_2
(1+\alpha_1)\;\;.\label{kklteta2}\eeq Requiring $\eta^{\rm max}\ll 1$ leads to
either $\alpha_1\ll 1$ or $\alpha_2\ll 1$.

The other three subcases are either uninteresting or equivalent to the former
case: If we change both the relative minus sign of $D_1$, $D_2$ and the
hierarchy of $\alpha_1$, $\alpha_2$ we are back to the former case with
exchanged labels ($1\leftrightarrow 2$). If we change just one of them we get a
$\delta V_2$ which has a global minimum with negative potential instead of the
desired maximum with positive potential.

In conclusion we cannot tune the maximum of the KKLT potential Eq.~\eqref{VWT2}
sufficiently flat by replacing its one additive uplift by a contribution of the
type of Eq.~\eqref{twoupl}.

\section{$T$-modulus inflation with $\alpha'$-corrections}\label{Tinfl}

The above result forces us to look for other minimal extensions of the setup
which may lead to saddle points with sufficiently small negative curvature. In
the literature~\cite{pill} a racetrack extension of the KKLT superpotential in
combination with an anti-D3-brane was used to construct an inflationary saddle
point.

We will show now that we can generate inflationary saddle points using the
following setup: the superpotential is given by \beq
W(T)=W_0+Ae^{-aT}+Be^{-bT}\;\;.\label{racet}\eeq Departing from~\cite{pill} the
uplift of the two degenerate $AdS$-minima present in the corresponding scalar
potential will now be provided by the $\alpha'$-corrected no-scale breaking
K\"ahler potential of Eq.~\eqref{dK} \beq K=-3\cdot\ln\left(T+\bar{T}\right)
-2\cdot\ln\left(1+\frac{\hat{\xi}}{2({2\;\textrm{Re}}\;T)^{3/2}}\right)\label{dK2}\eeq
This induces the contribution Eq.~\eqref{alphcorr} to the scalar potential. We
do not introduce an anti-D3-brane.

The analysis of the inflationary properties of the scalar potential given by
this setup follows closely the lines of~\cite{pill}. The differences (besides
using the $\alpha'$-correction instead of an anti-D3-brane) we will encounter
when looking at the structure of the minima and saddle points present in the
$\alpha'$-corrected scalar potential.

Assume now that the flux contribution $W_0$ has stabilized the dilaton $\tau$
in a minimum given by $\tilde{D}_{\tau}W=0$. Then the resulting scalar
potential can be written as \beq
V(T)=\left(1-\frac{\hat{\xi}}{({2\;\textrm{Re}}\;T)^{3/2}}\right)\,V_{\textrm{tree}}
+\frac{3}{8}\;e^{K^{(0)}}\frac{\hat{\xi}}{({2\;\textrm{Re}}\;T)^{3/2}}
\,\left|W\right|^2\label{alphcorr2}\eeq where \beq
K^{(0)}=-3\ln(T+\bar{T})\;\;.\eeq $V_{\textrm{tree}}$ denotes the scalar
potential induced by the above superpotential. It is given as \bea
V_{\textrm{tree}}(X,Y)&=&\frac{e^{-2(a+b)X}}{6X^2}\,\Big\{AB\left[3(a+b)+2abX\right]e^{(a+b)X}\cos[(a-b)Y]
\nonumber\\
&\phantom{=}&\hspace{2cm}
+aA\left[3\left(A+W_0\,e^{aX}\cos(aY)\right)+aAX\right]e^{2bX}\nonumber\\
&\phantom{=}&\hspace{2cm}
+bB\left[3\left(B+W_0\,e^{bX}\cos(bY)\right)+bBX\right]e^{2aX}\Big\}\;\;.\label{Vtree}\eea
Finally, $|W|^2$ reads \bea|W|^2&=&W_0^2+A^2e^{-2aX}+B^2e^{-2bX}+2A W_0
e^{-aX}\cos(aY)+2B W_0 e^{-bX}\cos(bY)\nonumber\\
&\phantom{=}&\hspace{0.5cm}+2A B e^{-(a+b)X}\cos[(a-b)Y]\;\;.\label{WW}\eea

\begin{figure}[ht]
\begin{center}
\includegraphics[width=12cm]{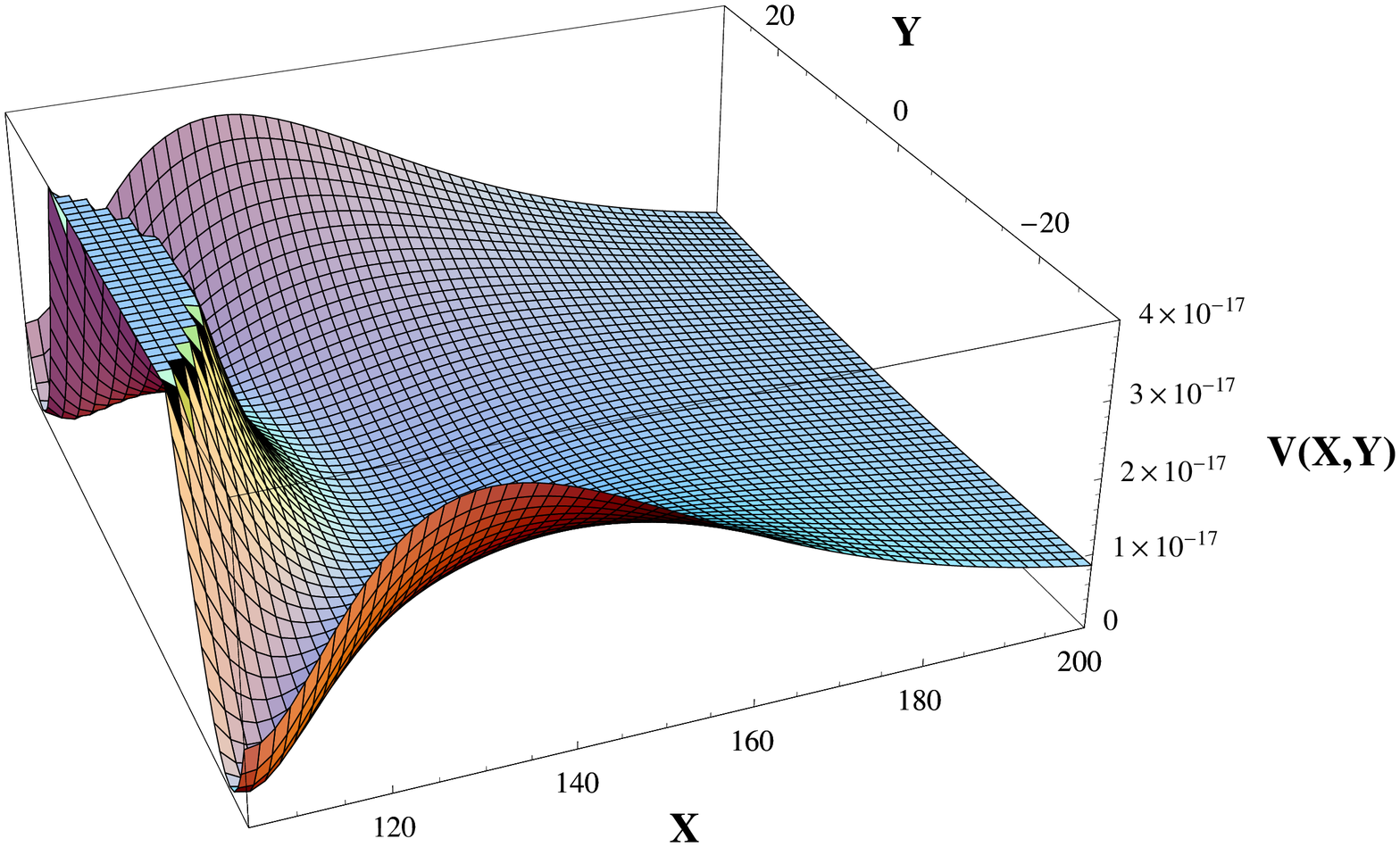}
\end{center}
\refstepcounter{figure}\label{saddle1}

\vspace*{-.2cm} {\bf Figure~\ref{saddle1}:} The scalar potential of $T$-modulus
with $\alpha'$-correction for a generic choice of parameters. Clearly visible
are the three minima connected by two off-$X$-axis saddle points.
\end{figure}
Compared to an anti-D3-brane uplift, the structure of this scalar potential is
changed considerably, since, as noted before, the $\alpha'$-uplift can only be
written locally as a purely additive contribution of the type $D/X^{\alpha}$.
Prior to uplifting we have a saddle at $Y=0$ which connects the two degenerate
$AdS$-minima at $Y_{\textrm{min}}^{(1)}=-Y_{\textrm{min}}^{(2)}\neq 0$ of the
scalar potential induced by the above superpotential. This saddle is rather
flat and extended in $X$ and $Y$. Therefore, unlike an anti-D3-brane uplift,
the $\alpha'$-contribution will not just lift the two minima to $V>0$ while
leaving the form of the saddle practically unchanged. The $\alpha'$-correction
will uplift and deform the initial saddle at $Y=0$ as well as it lifts the two
degenerate $AdS$-minima.

The shape of the potential arising this way looks the following: The initial
saddle point is at larger volume than the two $AdS$-minima and the
$\alpha'$-correction scales with an inverse power of the volume. Therefore, the
correction will raise the $AdS$-minima faster than the initial saddle point.
This implies that two new saddle points will appear which separate each of the
former $AdS$-minima from the region close to the former initial saddle point
which this way becomes a third local minimum. Therefore, after sufficient
uplifting we will have in general three different local minima at $V\geq 0$
with the properties \beq X_{\textrm{min}}^{(1)}=X_{\textrm{min}}^{(2)}\;,\;
Y_{\textrm{min}}^{(1)}=-Y_{\textrm{min}}^{(2)}\neq 0\;;\;
X_{\textrm{min}}^{(3)}>X_{\textrm{min}}^{(1)}\;,\;Y_{\textrm{min}}^{(3)}=0\;\;.\eeq
Two of them, $(1)$ and $(2)$, are each connected to the third one via a saddle
point. Fig.~\ref{saddle1} shows this situation for a generic choice of
parameters. The two saddle points have the properties
\[X_{\textrm{saddle}}^{(1)}=X_{\textrm{saddle}}^{(2)}=X_{\textrm{saddle}}\;,\;
Y_{\textrm{saddle}}^{(1)}=-Y_{\textrm{saddle}}^{(2)}\neq 0\] and furthermore
\beq X_{\textrm{min}}^{(1)}=X_{\textrm{min}}^{(2)}<X_{\textrm{saddle}}<
X_{\textrm{min}}^{(3)}\;\;.\eeq

This structure now allows for a new possibility of tuning the scalar potential
in order to find sufficiently flat saddle points: since the uplift of the
$\alpha'$-correction scales with a negative power of $X$, the two degenerate
minima $(1)$ and $(2)$ will get more strongly lifted than the saddle points
connecting them to minimum $(3)$ at $Y=0$. This third minimum, in turn, gets
even more weakly lifted than the saddle points. Hence, the potential can be
tuned in such a way that the minimum $(3)$ remains approximately Minkowski
while the two degenerate minima rise as a function of the uplift parameter
$\hat{\xi}$. Therefore, the saddle between minimum $(3)$ and, say, minimum
$(1)$ has very small negative curvature shortly before minimum $(1)$
disappears. The total set of parameters available ($A,B,a,b,W_0,\hat{\xi}$) is
large enough to allow for tuning both the curvature of these saddle points and
the vacuum energy $V(X_{\textrm{min}}^{(3)})$ of the approximate Minkowski
minimum $(3)$ to be small enough. For instance, imagine a situation where a
first tuning results in a situation with sufficiently small curvature of the
above two saddles and a hierarchy $0<V(X_{\textrm{min}}^{(3)})\ll V_{\rm
saddle}\sim V(X_{\textrm{min}}^{(1)})$. Then an additional fine-tuning of
$\hat{\xi}$ by a small amount $\delta\hat{\xi}$ allows for having
$V(X_{\textrm{min}}^{(3)})$ as close to zero as necessary to accomodate
$V(X_{\textrm{min}}^{(3)})\sim \Lambda_{\rm cosm.}$. This additional tuning
will not destroy the flatness of the saddle points since according to
Eq.~\eqref{alphaupl} the $\alpha'$-correction acts close to a given point, i.e.
a saddle point, similar to an additive anti-D3-brane uplift for a very small
change $|\delta\hat{\xi}|\ll\hat{\xi}$.

This mechanism is quite generic for a superpotential consisting of the flux
piece and two gaugino condensate contributions with its two degenerate
$AdS$-minima: it depends mainly on the hierarchy of the positions in $X$ of the
three minima and the two saddles that arise upon uplifting. Thus, even with
further $\alpha'$-corrections we expect this picture to remain qualitatively
the same, though the numerical values will change.

Firstly, we will show now that a considerable fine-tuning of $B$ is sufficient
to get enough $e$-foldings of slow-roll inflation on the saddle points. As an
example, consider the parameter choice \bea &&W_0=-5.55\cdot
10^{-5}\;,\;A=\frac{1}{50}\;,\;B=-3.37461131\cdot
10^{-2}\;,\;a=\frac{2\pi}{100}\;,\;
b=\frac{2\pi}{91}\nonumber\\
&&\hat{\xi}=-\,\frac{1}{2}\,\zeta(3)e^{-3\phi/2}\chi\;,\;\chi=-4209 \;\;.\eea
Here we assumed $e^{-3\phi/2}={\cal O}(1)$. Then the desired value of
$\hat{\xi}$ implies that we have to choose Calabi-Yau manifolds of large
negative Euler number with $\chi=-10^3\ldots-10^4$ which, in general, appears
to be possible~\cite{huschi}. For simplicity we set this quantity to unity
which leads to the above value of $\chi$.

Alternatively we can consider the possibility that the SM lives on a stack of
coincident D3-branes. The 4d gauge coupling on a stack of D3-branes is
$\alpha_{D3}=e^\phi /2$~\cite{Polch}. Phenomenologically
$\alpha_{\textrm{GUT}}=1/24$ and thus $e^{-\phi}\sim 12$ implying
$e^{-3\phi/2}\sim 50$. This reduces the absolute value of the Euler number
which is required to get the desired value of $\hat{\xi}$. As a numerical
example let us assume the dilaton fixed at $e^{-3\phi/2}=61$. Then we can
realize the above example for $\chi=-69$.
\begin{figure}[ht]
\begin{center}
\includegraphics[width=12cm]{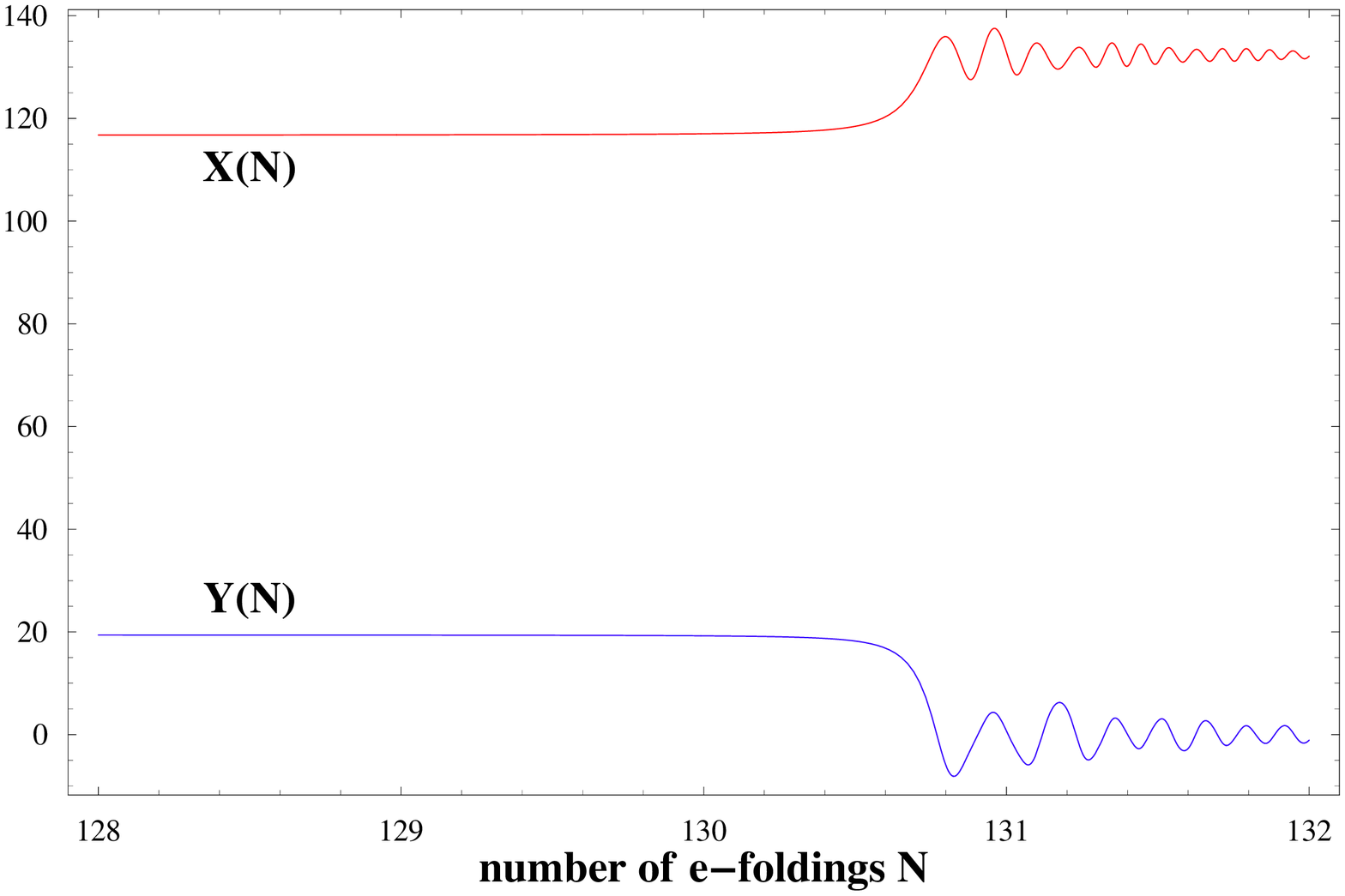}
\end{center}
\refstepcounter{figure}\label{infla2}

\vspace*{-.2cm} {\bf Figure~\ref{infla2}:} Evolution of the inflaton $T=X+iY$
as a function of time measured by the number of $e$-folding $N$.
\end{figure}
Thus, the model does not have to rely on the existence of Calabi-Yaus with
$\chi<-1000$. Otherwise, we may choose $|\chi|$ smaller which will move the
above structure of three minima towards smaller $X$-values. However, in the
following discussions we will set $e^{-3\phi/2}=1$ everywhere.

For the example given above we find the minimum $(3)$ at approximately \beq
X_{\textrm{min}}^{(3)}=132.398\;,\;Y_{\textrm{min}}^{(3)}=0\eeq being weakly de
Sitter. The other two degenerate minima reside at \beq
X_{\textrm{min}}^{(1)}=X_{\textrm{min}}^{(2)}=116.724\;,\;Y_{\textrm{min}}^{(1)/(2)}=\pm
19.431\;\;.\eeq The two saddle points we find very close by at \beq
X_{\textrm{saddle}}=X_{\textrm{saddle}}^{(1)}=X_{\textrm{saddle}}^{(2)}=116.728\;,\;
Y_{\textrm{saddle}}^{(1)/(2)}=\pm 19.428\;\;.\eeq As a consistency check we may
calculate the ratio \beq \frac{\hat{\xi}}{(2X)^{3/2}}\label{exppar}\eeq at the
three minima. This ratio is the expansion parameter used in deriving
Eq.~\eqref{alphcorr2} from Eq.~\eqref{dK2}. We find
$\hat{\xi}/(2X)^{3/2}\approx 0.5<1$ for the minimum $(3)$ and
$\hat{\xi}/(2X)^{3/2}\approx 0.7<1$ for the other two degenerate minima $(1)$
and $(2)$. This implies that the region around the three minima still resides
in the perturbative regime of the effective potential.

We may now calculate the Hesse matrix of curvatures \beq {\cal
H}=\left(\begin{array}{ccc}\frac{\partial^2V}{\partial X^2} &
\frac{\partial^2V}{\partial X\partial Y}\\ \frac{\partial^2V}{\partial
X\partial Y} & \frac{\partial^2V}{\partial Y^2}
\end{array}\right)\eeq diagonalize it and calculate from it the matrix of
slow-roll parameters on one of the saddle points to yield \beq {\cal
H}_{\eta}=\frac{2}{3}\,X_{\textrm{saddle}}^2{\cal H}^{\textrm{diag}}\approx
\left(\begin{array}{ccc}1222.83 & 0\\ 0 & -0.069
\end{array}\right)\;\;.\label{Hesse}\eeq Therefore, on these two saddle points,
slow-roll inflation can take place
if the $T$-modulus starts from the saddle with initial conditions fine-tuned to
some amount. For example, for initial conditions given by \beq
X_0=X_{\textrm{saddle}}^{(1)}+10^{-6}\;\;,\;\;Y_0=Y_{\textrm{saddle}}^{(1)}\;\;,\;\;
\dot{X}_0=\dot{Y}_0=0\eeq we get slow-roll inflation with some
$130\,e$-foldings and rolling-off into the $dS$-minimum $(3)$ of our world, as
seen in Fig.~\ref{infla2}. Here the equations of motion for the $T$-modulus
Eq.~\eqref{XYeom1} have been rewritten using
\begin{figure}[ht]
\begin{center}
\includegraphics[width=12cm]{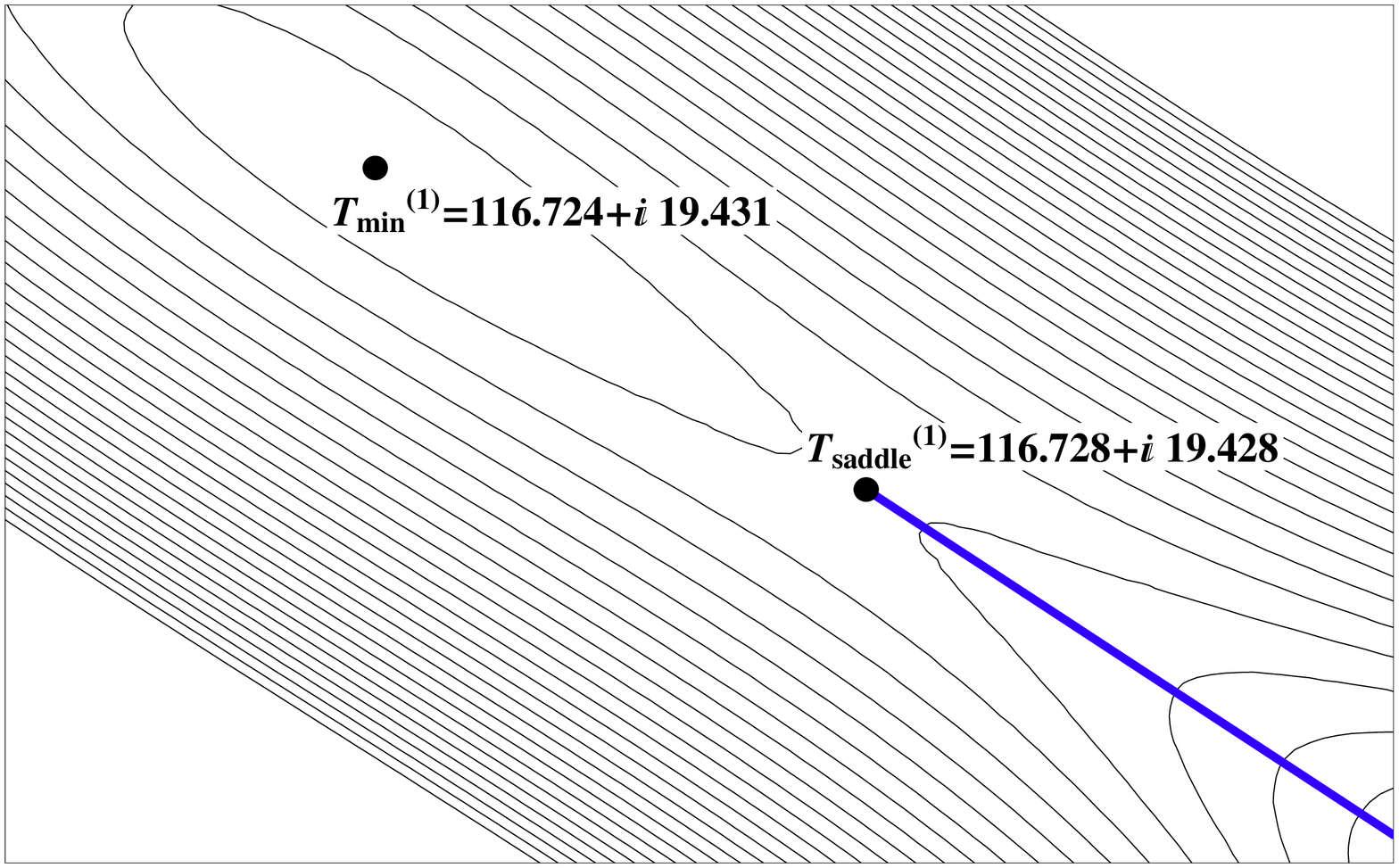}
\end{center}
\refstepcounter{figure}\label{saddle2}

\vspace*{-.2cm} {\bf Figure~\ref{saddle2}:} Contour plot of the potential close
to the saddle point $(1)$ and the evolution of the inflaton trajectory (thick
line)
in field space. The local minimum $(1)$ and the saddle point $(1)$ are indicated.
The contour lines curving away
from the starting point of the inflaton clearly indicate the saddle point
nature of this region. The long thin ellipse in the upper left encloses the
local minimum $(1)$.
\end{figure}
\bea \frac{\partial}{\partial
t}\,&=&H\,\frac{\partial}{\partial N}\;\;,\;\;\textrm{from the FRW scale factor
 }R(t)=e^{Ht}=e^N\nonumber\\
H^2&=&\frac{1}{3}\,\left[\frac{3}{4X^2}\,(\dot{X}^2+\dot{Y}^2)+V(X,Y)\right]\nonumber\\
&=&\frac{1}{3}\,V(X,Y)\cdot\left(1-\frac{X'^2+Y'^2}{4X^2}\right)^{-1} \eea to
yield~\cite{pill} \bea
X''&=&-\left(1-\frac{X'^2+Y'^2}{4X^2}\right)\left(3X'+2X^2\,\frac{1}{V}\,\frac{\partial
V}{\partial X}\right)+\,\frac{X'^2-Y'^2}{X}\nonumber\\
Y''&=&-\left(1-\frac{X'^2+Y'^2}{4X^2}\right)\left(3Y'+2X^2\,\frac{1}{V}\,\frac{\partial
V}{\partial Y}\right)+\,\frac{2X'Y'}{X}\label{XYeom2}\eea and $'$ denotes
$\partial/\partial N$. The structure of the potential and the initial part of
the inflaton trajectory in field space close to the saddle point can be found
in Fig.~\ref{saddle2}.

The Hubble parameter at the saddle point \beq
H_{\textrm{saddle}}=\sqrt{\frac{1}{3}\,V_{\textrm{saddle}}}\approx 10^{-9}\eeq
is much smaller than the initial fine-tuning of the inflaton on the saddle.
Thus, the scalar field fluctuations generated during inflation being of order
$H/2\pi={\cal O}(10^{-10})$ here~\cite{koltu} will not destroy the slow-roll
motion of the field.

We should mention here that by stronger fine-tuning in the potential the
slow-roll parameter $\eta$ of the saddle points can be made much smaller than
in the above numerical example. In this case, the amount of fine-tuning in the
initial conditions of the inflaton necessary to achieve sufficiently many
$e$-foldings can be relaxed. Thus, we may trade fine-tuning of the initial
conditions for fine-tuning of the potential.

Fine-tuning of the potential may be acceptable if we consider the extremely
large number of vacua the landscape contains. This large number allows us to
think of the parameters of the potential as being scanned sufficiently finely
across the landscape. In this view we also have no problem with the severe
fine-tuning already present in the potential. Since the potential arises from a
racetrack superpotential we naturally expect a fine-tuning in the parameters if
we balance the exponential contributions of the racetrack type to get flat
saddle points. In the above example the parameter $B$ was fine-tuned on the
level of about $10^{-8}$ which can be compared with the racetrack model
of~\cite{pill} where a fine-tuning of about $10^{-4}\ldots 10^{-3}$ was needed
to obtain sufficient slow-roll inflation. Since the number of vacua in the
string landscape is roughly $10^{500}$~\cite{dougl,sussk} we expect a much
higher level of fine-tuning allowed by the landscape.

Finally, we have the fact that within the string landscape the potential is
tuned discretely by the fluxes. We may consider this as an advantage compared
to purely field theoretic inflation models, where the potential can be
fine-tuned continuously in its parameters.

\section{Rescaling properties}\label{rescale}

The setup under discussion has certain scaling properties which are similar to
those of the scalar potential of~\cite{pill}. $|W|^2$ contains according to
Eq.~\eqref{WW} $a$, $b$, and $X,Y$ only in the combinations $aX$, $aY$, $bX$,
and $bY$ while in $V_{\rm tree}$ (see Eq.~\eqref{Vtree}) each term also has a
factor of either $a/X^2$ or $b/X^2$. Consider the rescaling \beq T\to \lambda
T\;,\; a\to \frac{a}{\lambda}\;,\;b\to
\frac{b}{\lambda}\;,\;\hat{\xi}\to\lambda^{3/2}\hat{\xi}\;\;\;\;\textrm{for:
}\lambda
>0\label{rescale1}\eeq where we leave the values of $W_0$, $A$ and $B$
unchanged. Then
the potential Eq.~\eqref{alphcorr2} itself rescales as \beq V\to
\frac{V}{\lambda^3}\;\;.\eeq Thus, the whole structure of the three minima and
two saddle points found above shifts along the $X$-axis. In the rescaled model
the stationary points reside at \bea
X'^{(i)}_{\textrm{saddle/min/max}}&=&\lambda\cdot
X_{\textrm{saddle/min/max}}^{(i)}\nonumber\\
Y'^{(i)}_{\textrm{saddle/min/max}}&=&\lambda\cdot
Y_{\textrm{saddle/min/max}}^{(i)}\eea respectively.  The eigenvalues of the
slow-roll parameter matrix ${\cal H}_{\eta}$ are invariant under this
rescaling. This is clear from Eq.s~\eqref{slow2} and \eqref{Hesse} since the
scaling $\partial_j\to \lambda^{-1}\partial_j$ ($j=X,Y$) implies
$(V'/V)^2\to\lambda^{-2}(V'/V)^2$ and $V''/V\to\lambda^{-2}V''/V$. Here $'$
denotes a derivative with respect to either $X$ or $Y$. The power spectrum of
density fluctuations generated during inflation \beq P_{\cal R}=
\frac{1}{24\pi^2}\,\frac{V}{\epsilon}\eeq scales upon the transformation
Eq.~\eqref{rescale1} as $P_{\cal R}\to \lambda^{-3}P_{\cal R}$.

Note further that $A$, $B$ and $W_0$ appear in both Eq.~\eqref{Vtree} and
\eqref{WW} only as polynomial products of degree two. A rescaling \bea &&T\to
\lambda T\;,\; a\to \frac{a}{\lambda}\;,\;b\to
\frac{b}{\lambda}\;,\nonumber\\
&&\hat{\xi}\to\lambda^{3/2}\hat{\xi}\;,\;A\to\lambda^{3/2}A\;,\;B\to\lambda^{3/2}B
\;,\;W_0\to\lambda^{3/2}W_0\;\;\;\;\textrm{for: }\lambda
>0\label{rescale2}\eea implies then that besides $\epsilon$ and $\eta$ also the
full scalar potential is invariant $V\to V$. Therefore, the transformation
Eq.~\eqref{rescale2} leaves the density fluctuation power spectrum unchanged.

We will rely heavily on these scaling properties of the model in the next
Section where we will search for a phenomenologically viable set of model
parameters.

\section{Experimental constraints and signatures}\label{ExpCon}

A realistic model of inflation has to generate a nearly scale-invariant power
spectrum of density fluctuations of the right magnitude. The fine-tuning of $B$
we chose in Section~\ref{Tinfl} was sufficient in order to obtain more than the
required 60 $e$-foldings of slow-roll inflation. In general this first step of
fine-tuning does not guarantee the density fluctuations at the COBE
normalization point at $N\approx 80$, i.e., about 55 $e$-foldings before the
end of inflation, to be small enough or to have a spectral index $n_s\approx
1$.

Therefore, we have to perform an additional fine-tuning: Using the rescaling
properties of the previous Section we have to shift the relevant part of the
scalar potential along the $X$-axis in order to search for a region where the
density fluctuations become small enough. And we need an additional fine-tuning
in $B$ to get saddle points with a slow-roll parameter $\eta$ small enough for
a viable $n_s$. By tuning of $B$ and the use of the rescalings given in the
Eq.s~\eqref{rescale1} and \eqref{rescale2} we find a new set of parameters \bea
&&W_0=-\,\frac{37}{46}\cdot10^{-6}\;,\;A=\frac{1}{3450}\;,\;B=-\,\frac{14672223067}{3\cdot
10^{13}}\nonumber\\&&a=\frac{2\pi}{100}\,\left(\frac{69}{10}\right)^{2/3}\;,\;
b=\frac{2\pi}{91}\,\left(\frac{69}{10}\right)^{2/3}\;,\;
\hat{\xi}=-\,\frac{1}{2}\,\zeta(3)\chi\;,\;\chi=-610 \;\;.\eea Here we have
assumed as before $e^{-3\phi/2}=1$ for simplicity.

This model contains again the two inflationary saddle points. However, their
negative curvature eigenvalue is now reduced and yields a slow-roll parameter
$\eta=-0.0064$. Solving the equations of motion for this rescaled model with
initial conditions given by \beq
X_0=X_{\textrm{saddle}}^{(1)}+\left(\frac{69}{10}\right)^{-2/3}\cdot 2.7\cdot
10^{-4}\;\;,\;\;Y_0=Y_{\textrm{saddle}}^{(1)}\;\;,\;\;
\dot{X}_0=\dot{Y}_0=0\eeq leads again to about 137 $e$-foldings of inflation
with the $X$ and $Y$ fields behaving very similar to the first case shown in
Fig.~\ref{infla2}.

Now calculate again the magnitude of the density fluctuations at the COBE
normalization point. The result at about 55 $e$-foldings before the end of
inflation corresponding to $N\approx 80$ is now \beq
\left(\frac{\delta\rho}{\rho}\right)_{k_0}\approx 2\cdot 10^{-5}\eeq yielding
the correct magnitude.
\begin{figure}[ht]
\begin{center}
\includegraphics[width=12cm]{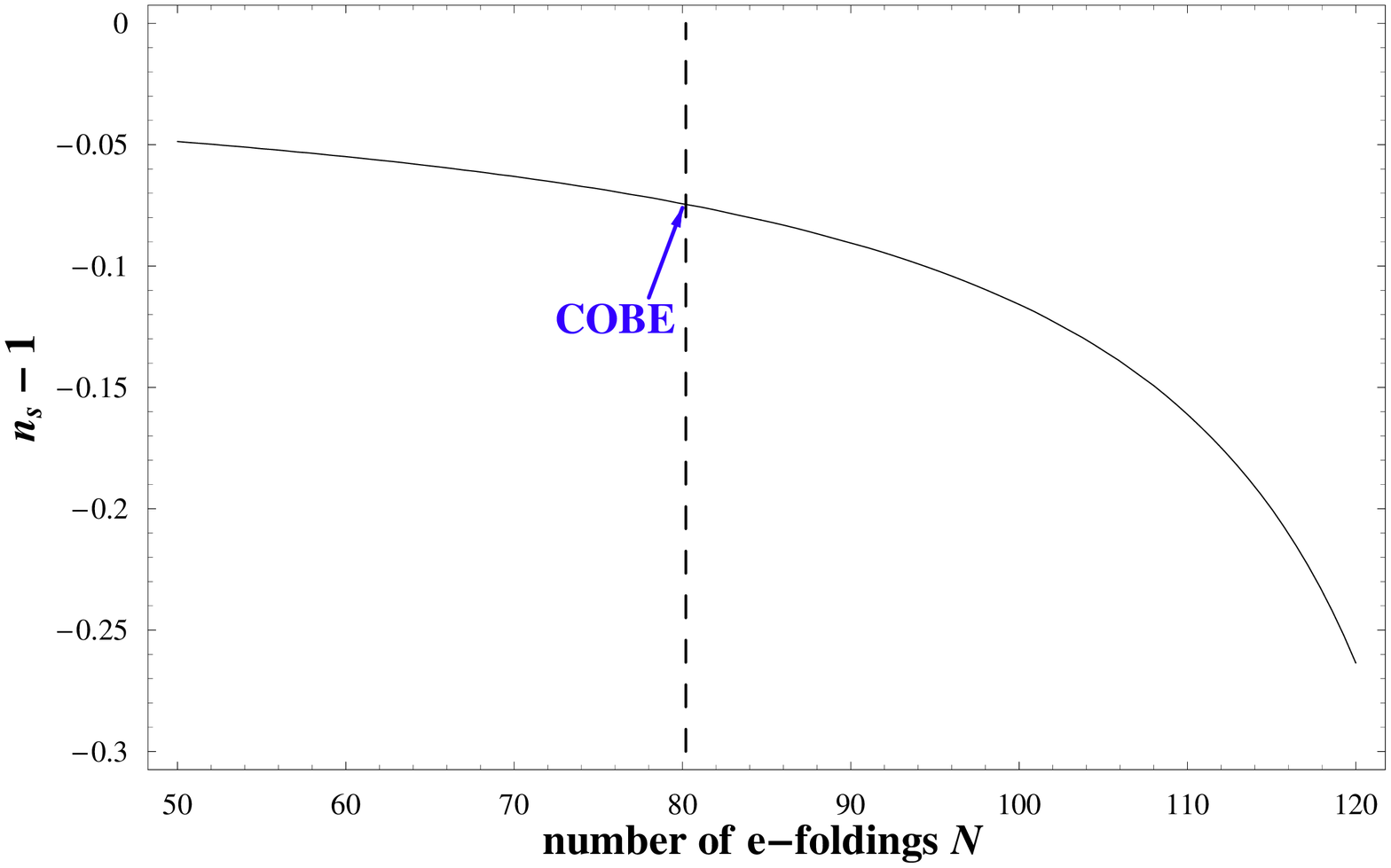}
\end{center}
\refstepcounter{figure}\label{infla4}

\vspace*{-.2cm} {\bf Figure~\ref{infla4}:} The deviation of the spectral index
from unity $n_s-1$ as a function of the number of $e$-foldings $N$. The COBE
normalization point sits at about 55 $e$-foldings before the end of inflation,
i.e., here at $N\approx 80$.
\end{figure}

Next, the spectral index is given by \beq
n_s=1+\left.\frac{d\ln {\cal P}_{\cal R}(k)}{d\ln
k}\,\right|_{k=RH}=1+2\,\left.\frac{d\ln( \delta\rho/\rho)}{d\ln
k}\,\right|_{k=RH}\label{specind2}\eeq evaluated as usual at horizon crossing.
Note that here we can replace $d\ln k\simeq dN$ because $k$ is evaluated at
horizon crossing $k=RH\sim H e^N$. Then we arrive at \beq n_s=1+2\,\frac{d\ln(
\delta\rho/\rho)}{dN}\eeq which results in the curve shown in
Fig.~\ref{infla4}.

The spectral index at the COBE normalization point therefore
yields a value of \beq n_s\approx0.93\eeq which is at $1\sigma$
consistent with the combined 3-year WMAP + SDSS galaxy survey
result $n_s=0.948^{+0.015}_{-0.018}$~\cite{WMAP} (the 3-year WMAP
data alone give $n_s=0.951^{+0.015}_{-0.019}$). However, the
numerical value of $n_s$ which we give here is a result of the
limited parameter space explored and does not imply a strict upper
bound on $n_s$ in the model. For comparison with the 3-year WMAP
data we give in addition the tensor-scalar ratio
$r=12.4\cdot\epsilon$ and the running spectral index $dn_s/d\ln
k=-16\epsilon\eta+24\epsilon^2+2\xi_{\rm infl}^2$ (where $\xi_{\rm
infl}^2=(2X^2/3)^2\cdot V'V'''/V^2$). We find at $N=80$ the values
$\epsilon\approx 3\cdot 10^{-14}$, $\eta\approx -0.035$ and
$\xi_{\rm infl}^2\approx -7\cdot 10^{-4}$. Thus, we have
negligible tensor contributions $r\approx 4\cdot 10^{-13}$ and
very small running $dn_s/d\ln k\approx -0.0014$.

Note that for the parameters chosen the rescaling places the post-inflationary
4d $dS$-minimum of our universe at $X_{\textrm{min}}^{(3)}=36.53$ and
$Y_{\textrm{min}}^{(3)}=0$. Thus, the 4d gauge coupling on a stack of D7-branes
in this $dS$ minimum is given by $\alpha\sim 1/X_{\textrm{min}}^{(3)}\approx
1/37$. This is not far from the phenomenological requirement $\alpha\sim 1/24$
allowing a construction of the Standard Model on a stack of intersecting
D7-branes. Therefore we have now both possibilities to place the Standard Model
on stacks of D3-branes or D7-branes.

\section{Eternal saddle point inflation}\label{infla3}

A check of the above numerical results is warranted. Therefore, we should study
the equations of motion Eq.~\eqref{XYeom1} of the non-canonically normalized
field $T$ in such KKLT-like setups in the vicinity of a saddle point. For
simplicity just concentrate on the equation of motion for the $X$-component.
Next assume that the saddle point at $X_s$ is tachyonic with negative curvature
in the $X$-direction. Then in its vicinity the potential can be approximated by
\beq V(X)=V_s-\frac{1}{2}\,|V_s''| (X-X_s)^2\;\;.\label{Vsaddle}\eeq Here $'$
denotes differentiation with respect to $X$. For a canonically normalized
scalar field the properties of inflation caused by the scalar field rolling
down from the saddle point have been studied in~\cite{LindFaRo}. Following the
lines of the analysis given there, we first rewrite the equation of motion for
$X$ in terms of the field $\phi=X-X_s$. The field will roll down from the
saddle into a local minimum with $|X_{\textrm{min}}-X_s|<<X_s$. Thus, $\phi$
obeys \beq
\ddot{\phi}+3H\dot{\phi}+\frac{1}{X_s}\,\dot{\phi}^2-\frac{2}{3}\,X_s^2
|V_s''|\phi=0\;\;.\eeq Using the ansatz \beq \phi(t)=\phi_0e^{\omega t}\eeq
this becomes \beq \omega^2+3H\omega+\frac{\omega^2\phi}{X_s}-\frac{2}{3}\,X_s^2
|V_s''|=0\;\;.\eeq Since we will analyze a regime where the Hubble parameter is
still dominated by the potential energy of $\phi$ and $\phi$ is very slowly
moving, one may assume $\omega^2\phi\ll X_s$. We will justify this in the end.
Now let us focus on the exponentially growing solution given by \bea \omega
&=&\frac{3}{2}\,H\left(-1+\sqrt{1+\frac{8}{9}\,X_s^2\frac{|V_s''|}{3H^2}}\right)\nonumber\\
&=&H\cdot |\eta_s|\eea where the slow-roll parameter is again defined as above
\beq|\eta_s|=\frac{2}{3}\,X_s^2\frac{|V_s''|}{V_s}\;\;.\eeq As a check of the
approximation made, we may plug in the simple example of KKLT discussed in
Sect.~\ref{nokklt}. We have from there $|\eta_s|=\frac{4}{3}\,aX_s$ and
$V_s\sim \frac{D}{X_s^2}$. Thus
\beq\omega=H|\eta_s|\sim\frac{4}{3\sqrt{3}}\,\frac{a\sqrt{D}}{X_s}\approx
10^{-9}\ll X_s\;,\;\;\textrm{for: }a\approx0.1\textrm{ and }X_s\approx
130\;\;D\approx 10^{-12}\eeq which satisfies the assumption $\omega^2\phi\ll
X_s$ a posteriori (the value of the field at the end of inflation is at most
$\phi_{\rm end}={\cal O}(10)$ in the KKLT example above).

Denoting now the value of field at the time where inflation ends with
$\phi^{\ast}$ we can derive the number of $e$-foldings in this fast-roll
inflation scheme as given by \beq
N=\frac{1}{|\eta_s|}\,\ln\left(\frac{\phi^{\ast}}{\phi_0}\right)\;\;.\label{Nefolds}\eeq
The final value $\phi^{\ast}$ here is determined either by the fact that the
potential and thus the Hubble constant have decreased significantly (this works
if the potential is very well described by the quadratic approximation even for
large $\phi$) or that at $\phi^{\ast}$ we have reached $|\eta|={\cal O}(10)$.
The last condition arises from Eq.~\eqref{Nefolds}. For $|\eta|=6\ldots10$ even
a very large ratio $\phi^{\ast}/\phi_0\sim M_p/M_{EW}\sim 10^{17}$ does not
generate more than about $10$ additional $e$-foldings.

As a check of the numerical results of the last Section we may apply now these
results. The number of $e$-foldings there is given by Eq.~\eqref{Nefolds} in
terms of the initial deviation of the inflaton field from the saddle point
$\phi_0$, the final value $\phi^{\ast}$ when inflation ends and the saddle
curvature in its tachyonic direction $\eta_s$ as \beq
N=\frac{1}{|\eta_s|}\,\ln\left(\frac{\phi^{\ast}}{\phi_0}\right)\;\;.\eeq Now
in the first example of the last Section $\phi_0=10^{-6}$ (see above). Further,
we have $|\eta_s|=0.069$. It remains to determine $\phi^{\ast}$ as the end
point of the inflationary phase. For this purpose we have to analyze the
potential $V(X(N),Y(N))$ along the inflationary trajectory above and to
calculate the $\eta$-values along the trajectory. We find that when the
$T$-modulus has moved to a distance of about $0.01$ from the saddle,
$\eta\approx -10$ which means that inflation effectively ends there. Plugging
this now in the above formula we obtain \beq
N=\frac{1}{|\eta_s|}\,\ln\left(\frac{\phi^{\ast}}{\phi_0}\right)\approx
133\;\;.\eeq This is sufficiently close to the purely numerical results above,
which indicates that the numerical solution is stable and closely resembles the
true one.

Note that in this model each of the two rather flat saddle points still
connects two minima ($(1)$ and $(3)$ or $(2)$ and $(3)$, respectively). In such
a situation, where a sufficiently flat saddle point connects two minima along a
certain direction in field space, inflation may also arise from inflating
topological defects, namely, domain walls~\cite{Vil}. It is therefore tempting
to speculate that besides slow-roll inflation also eternal topological
inflation arises on the saddles constructed here, which would relieve the
question of fine-tuning the initial conditions of the
inflaton~\cite{lind1,pill}. The original literature~\cite{Vil,lind1} uses a
saddle point connecting two degenerate minima in deriving the conditions for
topological inflation: the saddle curvature has to be small enough that
$\eta_{\textrm{saddle}}\ll 1$, which corresponds to domain walls whose wall
thickness is large compared to their gravitational radius. As an illustration
consider the example of static domain walls of the $Z_2$-symmetric theory \beq
{\cal L}=\frac{1}{2}\,(\partial_\mu\phi)^2-V(\phi)\;\;,\;
V(\phi)=\frac{\lambda}{4}\,(\phi^2-\beta^2)^2\eeq which are given by the
solution \beq \phi_{\rm wall}(x)=\beta\tanh\Big(\sqrt{\frac{\lambda}{2}}\,\beta
x\Big)\eeq for a wall in the $yz$ plane. The thickness of the wall $\delta$ is
determined by the equilibrium of gradient and potential energy density as \beq
\left.\rho_{\rm grad}\right|_{x\sim\delta}\sim\frac{\beta^2}{\delta^2}\sim
\rho_{\rm pot}=V(0)\sim \lambda\beta^4
\;\Rightarrow\;\delta\sim\frac{1}{\beta\sqrt{\lambda}}\;\;.\eeq
The gravitational radius of the wall is $R=2M_{\rm wall}\sim 8\pi\rho\delta^3/3$
where the energy density is $\rho=\lambda\beta^4/2$ (the sum of the potential
energy density and the gradient energy density).
\begin{figure}[ht]
\begin{center}
\includegraphics[width=12cm]{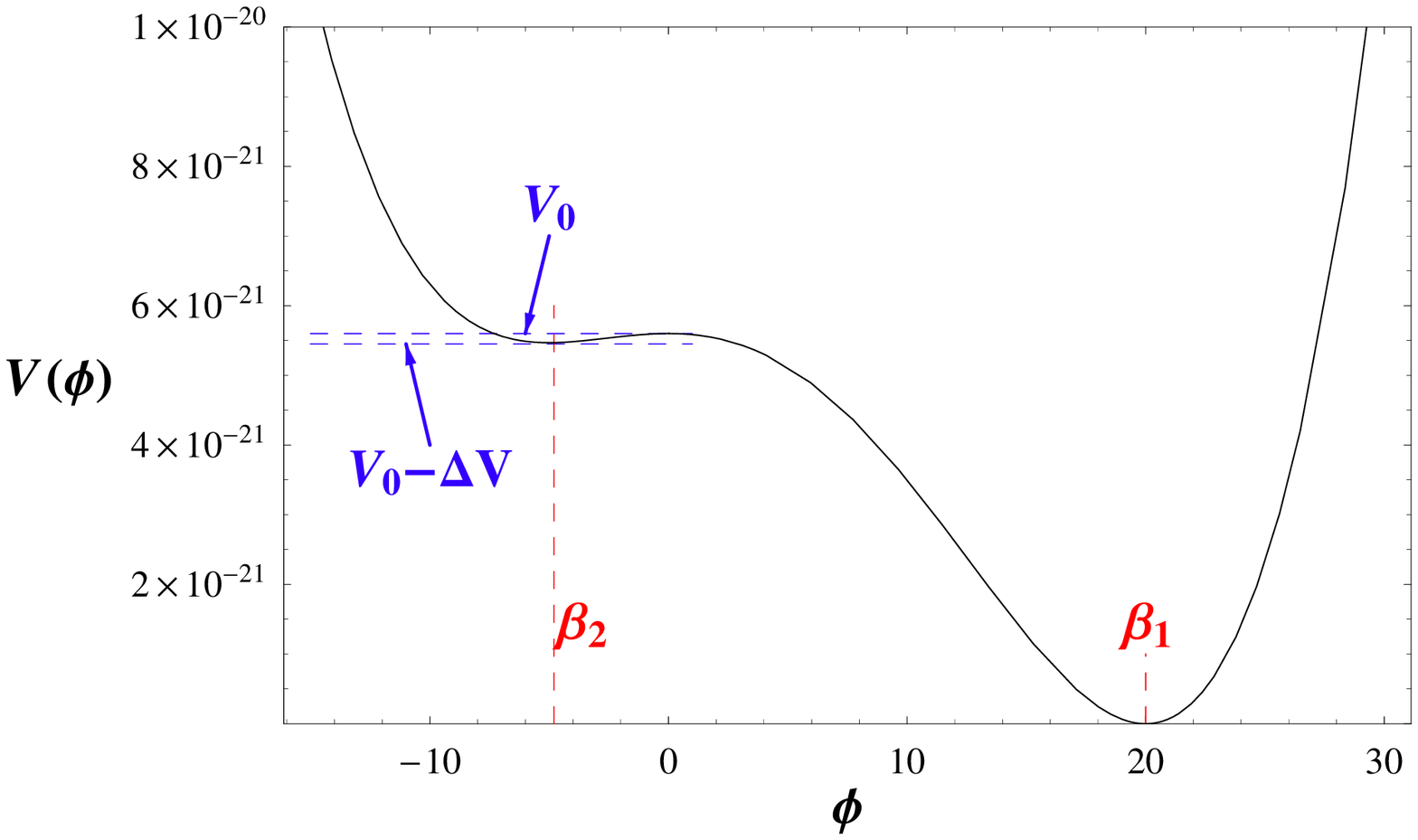}
\end{center}
\refstepcounter{figure}\label{potasymm}

\vspace*{-.2cm} {\bf Figure~\ref{potasymm}:} A highly asymmetric double-well
scalar potential as it is realized along the inflaton trajectories in the
previous Section. $\Delta V$ and $\beta_2$ are exaggerated compared to the
values in the actual model.
\end{figure}
Gravitational effects become
important once the gravitational radius exceeds the wall thickness, i.e, for
\beq \delta<R\;\Rightarrow\;\beta>\frac{3}{4\pi}\label{topinfcond}\eeq in
Planck units. If we calculate the slow-roll parameter $\eta$ at the center of
the wall the result is $\eta_{x=0}=V''(0)/V(0)=4/\beta^2$. Requiring $\eta<1$
therefore corresponds to the previous 'importance of gravity' condition. The
above static wall solution would never inflate since the potential and gradient
energy density are of the same order near the wall. However, if inflation
started in a small patch of space-time with $\phi=0$ then the fluctuations
$\delta\phi\sim H$ with wavelength $H^{-1}$ generated after each time interval
$H^{-1}$ have a gradient energy $\sim H^4\sim V^2<<V$ as long as $V<<1$ in the
wall. In this case, an initially inflating wall which fulfills
Eq.~\ref{topinfcond} will continue to inflate forever near to the wall
center~\cite{Vil,lind1}.

This analysis is valid in the symmetric potential of the example above. In the
cases under consideration in the last Section the saddles connect two highly
non-degenerate minima \beq \frac{V_{\textrm{saddle}}^{(1)}-
V_{\textrm{min}}^{(1)}}{V_{\textrm{saddle}}^{(1)}} =\,\frac{\Delta
V}{V_{\textrm{saddle}}^{(1)}}\ll \frac{V_{\textrm{saddle}}^{(1)}-
V_{\textrm{min}}^{(3)}}{V_{\textrm{saddle}}^{(1)}}\approx 1\;\;.\eeq
Fortunately, we can extend the above analysis to this situation. For simplicity
we will render the problem again one-dimensional. This is possible by looking
at the scalar potential along the inflation trajectories of the examples of the
previous Section. This effectively one-dimensional potential then looks like
the one shown in Fig.~\ref{potasymm}.

In this potential, too, there will be a domain wall like solution $\phi_{\rm
wall}$ with the properties \beq \phi_{\rm wall}:
\left\{\begin{array}{c}\lim_{x\to-\infty}\phi_{\rm wall}= \beta_2<0\\
\lim_{x\to\infty}\phi_{\rm wall}=\beta_1>0\\
\phi_{\rm wall}(x=0)=0
\end{array}\right.\;\;.\label{wall1}\eeq
This solution is no longer symmetric under $x\to -x$. In particular, it can be
described by two wall thickness parameters $\delta_1$ and $\delta_2$ for $x>0$
and $x<0$, respectively. For $x<0$ the gradient energy of the wall has to
compensate just the small potential energy difference $\Delta V$ between the
$\phi$-maximum and the minimum at $\phi=\beta_2<0$. The gradient energy at
$x>0$, however, compensates for the full potential $V_0$ of the $\phi$-maximum.
Thus, we get from the equilibrium of potential and gradient energy the
relations \beq \delta_1\sim \frac{\beta_1}{\sqrt{V_0}}\;\;,\;\; \delta_2\sim
\frac{|\beta_2|}{\sqrt{\Delta V}}\;\;.\eeq The wall becomes dominated by
gravity if $\delta_1+\delta_2<R\sim\rho(\delta_1+\delta_2)^3$ which results in
a condition \beq \delta_1+\delta_2>\,\frac{1}{\sqrt{V_0}}\sim
H^{-1}\;\;.\label{topinfcond2}\eeq For, e.g., $\delta_2>\delta_1$ this is
essentially Eq.~\eqref{topinfcond}. If in addition $V_0\ll 1$ holds, a single
patch of size $\sim H^{-1}$, which is filled initially with a field
$\phi\approx 0$ with fluctuations $\delta\phi\ll H$, will become the 4d $dS$
core of an exponentially expanding wall as noted above already.

Note that the high-lying minimum also gives rise to a fast expanding $dS$
space-time. However, since the potential energy of the maximum always exceeds
the high-lying minimum, the space-time in the core of the wall with the field
on the maximum will expand faster than that of the high-lying minimum.

Once the field starts to roll down towards the post-inflationary minimum (3)
with a very small cosmological constant $V_{\rm min}^{(3)}\approx 0$ a bubble
of the new vacuum given by the minimum (3) is formed. Even without gravity the
bubble would expand since the energy density of the vacuum inside the bubble is
smaller than outside the bubble where it is given by the minimum (1) on the
other side of the saddle point~\cite{Coleman}. For this thin-wall case without
gravity the bubble wall would still be given by a kink solution of the form of
Eq.~\eqref{wall1}. However, the wall position would now be given by $x=0=R-R_0$
with $R=\sqrt{|\vec{r}(t)|^2-c^2 t^2}$ which describes a bubble wall which
expands with nearly the speed of light shortly after it is born~\cite{Coleman}.

Without gravity, this expanding bubble would finally convert all space-time in
the vacuum state of the minimum (1) to the one of minimum (3). However, as we
consider the case of a thick wall dominated by gravity which possesses a fast
inflating core, this over-roll of the outside space-time in the vacuum state of
the minimum (1) cannot happen. This is due to the fact that the core of the
wall expands exponentially fast. While the interior side of the wall of the
bubble of the vacuum (3) when viewed from inside recedes with nearly the speed
of light its outer side recedes exponentially fast. Therefore, the interior
side of the wall recedes exponentially fast from its outer side, and the bubble
can never convert all of the outside space-time into the vacuum inside the
bubble. The processes inside the wall are decoupled from the physics inside and
outside the bubble due to the de Sitter horizon formed by the exponential
expansion of the wall's core. Thus, once the appropriate conditions are
satisfied, eternal topological inflation may take place inside a thick wall
dominated by gravity even if the wall forms an expanding bubble due to
non-degenerate minima of the potential.

Now we concentrate on the quantum fluctuations of the scalar field $\phi$
inside the inflating core of the wall. For inflation to get started repeatedly
within the wall there must be a region close to the $\phi$-maximum where the
$dS$ quantum fluctuations of $\phi$ dominate its classical
evolution~\cite{Vil,lind1}. Initially we have $\ddot{\phi}\approx 0$ and thus
the slow-roll equation of motion of the non-canonically normalized field $\phi$
governs the classical dynamics close to the $\phi$-maximum \beq
\dot{\phi}=-\,\frac{2X_s^2}{3}\,\frac{V'(\phi)}{3H}\;\;.\eeq Now close the
$\phi$-maximum we can use Eq.~\eqref{Vsaddle} to arrive at \beq
\dot{\phi}=H\eta_s\phi\;\;.\eeq Within the time interval $\Delta t=H^{-1}$ the
field moves classically by \beq \delta\phi_{\rm class}=\eta_s\phi\;\;.\eeq
Simultaneously it receives a contribution from quantum fluctuations \beq
\delta\phi_{\rm quant}\sim H\;\;.\eeq The quantum fluctuations dominate the
classical motion (which drives the field down into the minima) for \beq
\phi<\phi^\ast\;\;{\rm with:}\;\;\phi^\ast\sim\,\frac{H}{\eta_s}\;\;.\eeq If
now $\phi^\ast\gg \delta\phi_{\rm quant}\sim H$ there is a region close to
$\phi=0$ at the center of the wall where the $dS$ quantum fluctuations of
$\phi$ can jump the field many times before eventually passing $\phi^\ast$ from
where the field moves classically. Therefore, within this region the field will
jump over and again arbitrarily close to $\phi=0$ thus starting inflationary
patches without end. Plugging in $\phi^\ast$ in $\phi^\ast\gg \delta\phi_{\rm
quant}\sim H$ leads to the condition \beq \eta_s\ll 1\eeq the slow-roll
condition.

Therefore, a highly asymmetric double-well potential shows eternal topological
inflation provided that 1) the slow-roll conditions hold on the maximum, 2) on
the maximum $V_0\ll 1$, and 3) the 'gravity domination' condition
Eq.~\eqref{topinfcond2} holds. We apply these conditions now to the realistic
example (the $2^{\rm nd}$ one) of the previous Section. There we have
$V_0={\cal O}(10^{-20})\ll 1$ and $\eta_s=0.0064\ll 1$. In terms of the above
notation we have further $\beta_2\sim-10^{-4}$, $\beta_1\sim 20$, and $\Delta
V\sim 10^{-14}\,V_0$. This implies \beq \delta_1\sim
10^{11}>\frac{1}{\sqrt{V_0}}\;\;,\;\;\delta_2\sim
10^{13}>\frac{1}{\sqrt{V_0}}\eeq which satisfies Eq.~\eqref{topinfcond2}.
Therefore, the inflation model of the previous Section has the property of
eternal topological inflation on its saddle points.

The initial probability of creating space-time regions where $T$ is close to
the saddle points of its potential is exponentially small. However, the
inflationary regions, which are seeded by eternal topological inflation,
dominate the volume of 4d space-time after inflation because of the exponential
growth. Therefore, the post-inflationary volume fraction of the universe which
is in the vacuum given by the 4d $dS$ minimum of $T$-modulus will be
large~\cite{Vil2}. This resolves the problem of fine-tuning the initial
conditions for the slow-roll inflationary phase which we otherwise would have
in the model of the previous Section~\cite{pill} (see also the recent
discussion in~\cite{hall}).

As a last comment, we note that the cosmological overshoot
problem~\cite{BruSteinHardt} as well as the problem of moduli destabilization
at high temperatures~\cite{BuchmDestab1} under certain conditions are absent in
our model. In order to see this look at the final 4d $dS$ minimum of presumably
our world at $X_{\rm min}^{(3)}$ (see the previous Sect. for the notation). If
our universe originated via eternal topological inflation on one of the saddle
points of the scalar potential at, e.g., $X_{\rm saddle}^{(1)}$ then the
reheating temperature after rolling down into the 4d $dS$ minimum at $X_{\rm
min}^{(3)}$ cannot exceed \beq T_{\rm reh}^{\rm max}\sim (V_{\rm
saddle}^{(1)})^{1/4}\;\;.\eeq The post-inflationary minimum at $X_{\rm
min}^{(3)}$, however, is separated from $X\to\infty$ by a maximum in $X$. The
potential of this maximum $V_{\rm barrier}$ in our model is given by\beq V_{\rm
barrier}\sim 3 V_{\rm saddle}^{(1)}\;\;.\eeq Thus, neither reheating nor the
kinetic energy of the $T$-modulus rolling down from the saddle point can drive
the field over the barrier.

\section{Conclusion}\label{con}

In this paper we analyze phenomenological aspects of higher-order
$\alpha'$-corrections in the context of moduli stabilizing flux
compactifications of the type IIB superstring. We discuss the
inflationary properties of the volume modulus in the original KKLT
setup. In the simplest class of these models - consisting of the
flux superpotential, the contribution of one gaugino condensate on
a stack of D7-branes, and a single additive uplifting potential of
a general inverse power-law form - slow-roll inflation ending in
the KKLT $dS$-minimum cannot occur. We study $\alpha'$-corrections
which are higher-order curvature corrections and thus
higher-dimension operators appearing in the K\"ahler potential of
the effective action. We demonstrate that the generic ability of
these higher-dimension operators to lift stable $AdS_4$ type IIB
string vacua to the desired metastable $dS$-minima for the $T$
modulus (the volume modulus) can also be used to provide slow-roll
inflation using the same $T$ modulus. Such a setup has no
$\eta$-problem because the leading order K\"ahler potential for
the $T$ modulus is of the no-scale type. We construct a concrete
model using fluxes and a racetrack superpotential which upon
inclusion of the $\alpha'$-corrections yields $T$-modulus
inflation on saddle points of the potential with some $130$
$e$-foldings. At the end of inflation the $T$-modulus rolls from
the saddle point down into a $dS$-minimum with a small positive
cosmological constant where the modulus is stabilized. The model
has certain scaling properties allowing us to shift the
inflationary region of the potential to different values of the
real part of $T$ while leaving the inflationary properties of the
saddle points invariant. We argue that these saddle points might
be generically present if racetrack superpotentials and
$\alpha'$-corrections are both taken into account. The model can
accommodate the 3-year WMAP data of the CMB radiation. It yields
primordial density fluctuations of the right magnitude with a
spectral index of these fluctuations $n_s\approx 0.93$. We point
out that eternal topological inflation occurs in the model which
removes the fine-tuning problem of inflationary initial
conditions. Finally, we comment on the cosmological overshoot
problem and the destabilization of the moduli at high
temperatures. These effects are absent in the fraction of the
universe which is seeded by topological eternal inflation in our
model.

\vspace*{1cm} \noindent {\bf Acknowledgements}: I would like to thank
W.~Buchm\"uller, A.~Hebecker, J.~Louis, A.~Lucas, V.~Rubakov and M.~Trapletti
for useful discussions and comments.


\end{document}